# First-Principles DFT Study of Ferroelectric-to-Antiferroelectric Phase Transitions in $LiTaO_3$ under Electric Boundary Conditions

Eva Greene and Shaohui Qiu
*Department of Chemistry and Physics, Southern Utah University, Cedar City, Utah 84720, USA*

## Abstract

Ferroelectric (FE) materials exhibit switchable polarization that enables non-volatile memory, while antiferroelectric (AFE) materials exhibit antiparallel dipoles that enable high-energy-density capacitors. We use first-principles density functional theory to investigate how electric boundary conditions select among competing structural phases in lithium tantalate ($LiTaO_3$). Phonon analysis of the high-symmetry reference structure reveals two relevant unstable modes: a polar $A_{2u}$ mode at $181i$ cm$^{-1}$ driving the FE phase, and an antipolar $A_{2g}$ mode at $103i$ cm$^{-1}$ driving the AFE phase. Under short-circuit boundary conditions, the FE phase is the global minimum with well depth ≈144 meV. Under open-circuit boundary conditions, the FE phase is destabilized and the AFE phase becomes the ground state with well depth ≈19 meV. A weakly unstable longitudinal-optic mode at $48.3i$ cm$^{-1}$ produces a well depth of only ≈0.24 meV, too shallow to be thermodynamically relevant. Mapping the free energy as a function of electric displacement $D$ identifies a critical value $D_c$ = 0.051 C/m² at which a second-order FE↔AFE transition occurs—lower than the corresponding 0.07 C/m² in $LiNbO_3$, making the transition more experimentally accessible.

## 1. Introduction

### *1.1 Ferroelectric materials and their applications*

Ferroelectric (FE) materials are characterized by a spontaneous electric polarization that can be reversed by an applied electric field. This polarization arises when the crystal structure breaks centrosymmetry below the Curie temperature, producing two equivalent polar configurations that correspond to opposite orientations of the net dipole moment. When polarization is plotted against free energy, FE materials exhibit a characteristic double-well landscape with two equally stable minima [3]. The ability to switch between these minima with an applied field, combined with the stability of each minimum once written, makes FE materials valuable for non-volatile memory devices, where the two polarization states encode binary 0 and 1 [4]. Ferroelectric random-access memory (FeRAM) is now commercially deployed in smart cards, RFID tags, automotive electronics, and medical monitors, taking advantage of fast read/write speeds, low power consumption, and high endurance.

FE materials also play important roles in pyroelectric infrared sensors, piezoelectric actuators and transducers, and energy-harvesting devices that convert mechanical or thermal energy into electrical energy through changes in polarization [3].

### *1.2 Antiferroelectric materials and their applications*

Antiferroelectric (AFE) materials, by contrast, exhibit an antiparallel arrangement of dipoles in adjacent unit cells, yielding zero net polarization in the ground state [5,6]. Under an applied electric field, AFE materials undergo a transition to a polar (FE-like) state, producing a characteristic double hysteresis loop in the polarization–field response. This field-induced AFE→FE transition stores energy efficiently because the high maximum polarization is combined with a low remnant polarization, resulting in low energy loss per cycle and fast discharge rates [7,8]. Recent demonstrations have achieved energy densities exceeding 20 J/cm$^3$ with energy-storage efficiencies above 90% in lead-free AFE ceramics, making AFE materials leading candidates for high-energy-density pulsed-power capacitors [9].

### *1.3 Boundary conditions and the depolarization field*

In a polarized FE material, surface bound charges produce a depolarization field that opposes the polarization itself. Whether this depolarization field is screened or unscreened depends on the electrical boundary conditions of the sample. Under short-circuit boundary conditions (SCBC), conducting electrodes provide free charges that compensate the bound surface charges, and the depolarization field vanishes; the FE phase is then stabilized. Under open-circuit boundary conditions (OCBC), no such screening occurs, the depolarization field reaches its maximum magnitude, and the FE phase is typically destabilized in favor of a paraelectric (centrosymmetric) phase [17,18]. Real thin films sit between these two limits depending on the quality of the electrodes [19].

These boundary conditions can be expressed in terms of the electric displacement field $D$. From Gauss's law, $\nabla \cdot D = \rho_{\text{free}}$, and under OCBC with no free charge available, $D = 0$. Combined with $D = \varepsilon_0 E + P$, the OCBC condition implies a strong internal field $E = -P/\varepsilon_0$ (modulo screening) that destabilizes the polarized state. SCBC corresponds to $E = 0$ with no constraint on $D$. Treating $D$ as the independent variable rather than $E$ provides a controlled way to interpolate between these two limits in first-principles calculations [20].

### *1.4 $LiTaO_3$ as a model system*

Lithium tantalate ($LiTaO_3$) and lithium niobate ($LiNbO_3$) are isomorphous ferroelectrics that crystallize in the rhombohedral $R3c$ space group below their Curie temperatures of approximately 950 K and 1480 K, respectively [10–12]. They exhibit large spontaneous polarizations of ≈60 $\mu C/cm^2$ ($LiTaO_3$) and ≈71 $\mu C/cm^2$ ($LiNbO_3$) at low temperatures [16]. Both materials are widely used in electro-optic modulators, surface-acoustic-wave devices, frequency doublers, and pyroelectric detectors [10].

Although structurally similar, $LiTaO_3$ has emerged in recent years as a particularly attractive platform for integrated photonics. It exhibits much lower optical birefringence than $LiNbO_3$, which enables high-density circuits and broadband operation across the entire telecommunications band [14]. It also offers a higher photorefractive damage threshold and lower microwave loss tangent, while retaining the strong electro-optic, piezoelectric, and pyroelectric responses that have made the $LiNbO_3$ family technologically important for decades [15].

### *1.5 Motivation and scope*

Conventional wisdom holds that an FE material under OCBC transitions to a paraelectric phase, severing any potential connection to AFE physics. However, recent first-principles work on $LiNbO_3$ has shown that the situation is more nuanced. Li and co-workers had earlier proposed, on the basis of an unstable LO phonon, that $LiNbO_3$ should be hyperferroelectric — capable of maintaining nonzero polarization under OCBC despite the strong depolarization field [21,22]. Subsequently, Qiu *et al.* [23] showed that the ground state of $LiNbO_3$ under OCBC is in fact nonpolar, driven by an antiferroelectric-like soft $A_{2g}$ mode at $120i$ $cm^{-1}$ — revealing that the presence of a soft LO mode does not, by itself, guarantee hyperferroelectric behavior. Building on this finding, Qiu and Fu [24] mapped the complete free-energy landscape of $LiNbO_3$ as a function of the imposed electric displacement and identified a pronounced second-order phase transition between the nonpolar (antiferroelectric-like) phase and the polar TO phase at a critical displacement of $D_c = 0.07$ $C/m^2$. Importantly, the original hyperferroelectric proposal of Li *et al.* [22] also extended to $LiTaO_3$, which possesses a similarly unstable LO mode and was likewise predicted to maintain nonzero polarization under OCBC. Whether the $LiNbO_3$ picture — nonpolar antiferroelectric-like ground state — or the original hyperferroelectric picture better describes $LiTaO_3$ has not previously been investigated by direct first-principles calculation.

The present work addresses this gap. We use first-principles density functional theory to identify the soft phonon modes of $LiTaO_3$, evaluate the free-energy landscapes of the resulting structural phases as functions of the electric displacement $D$, and quantitatively determine which phase is the ground state in each regime. We construct the complete $D$-field phase diagram of $LiTaO_3$ and compare the results in detail with the corresponding $LiNbO_3$ picture.

## 2. Methods

### *2.1 Density functional theory framework*

All electronic-structure calculations are carried out within density functional theory [1,2] using the local density approximation as implemented in the Quantum ESPRESSO package [28]. We use Troullier–Martins norm-conserving pseudopotentials [29] to represent the effect of core electrons. The Brillouin zone is sampled with a 4×4×4 Monkhorst–Pack mesh, and the plane-wave kinetic-energy cutoff is 110 Ry. Forces are converged to $5\times10^{-5}$ Ry/Bohr and energies to $10^{-4}$ Ry. The 10-atom rhombohedral unit cell is fully relaxed for each structural configuration considered.

### *2.2 Three complementary computational approaches*

Three methods are used in combination. First, the phonon frequencies and eigenvectors of the high-symmetry reference structure are computed using density functional perturbation theory (DFPT) [25]. Imaginary frequencies indicate dynamical instabilities and identify the candidate distortions that lead to lower-symmetry structural phases. We focus on zone-center modes since these correspond to homogeneous structural distortions of the unit cell.

Second, electric polarizations of the various distorted configurations are evaluated using the modern theory of polarization via the geometric Berry-phase approach [26,27]. The polarization $P(\lambda)$ of a configuration parameterized by displacement amplitude $\lambda$ is needed to evaluate the depolarization energy at each $D$ field.

Third, the free energy under a constrained electric displacement field is obtained following the Legendre-transformation approach of Stengel, Spaldin, and Vanderbilt [20]. For a given soft-mode eigenvector $u_i$, atoms are displaced from the centrosymmetric reference position $r^c_i$ according to $r_i(\lambda) = r^c_i + \lambda a_0 u_i$, where $a_0$ is the lattice constant. The free energy at this configuration under a finite $D$ field is

$$\Phi(\lambda) = U_{KS}(\lambda) + \frac{\Omega(\lambda)}{2\epsilon_0[1+\chi_\infty(\lambda)]} \cdot [D - P(\lambda)]^2 \quad (1)$$

where $U_{KS}$ is the Kohn–Sham internal energy, $\Omega$ is the unit-cell volume, and $\chi_\infty$ is the high-frequency dielectric susceptibility. The second term is the depolarization energy, denoted $U_{dp}$, and represents the energetic cost of mismatch between the imposed displacement field $D$ and the polarization $P$ of the configuration. Under SCBC the free energy reduces to $U_{KS}$ alone (no depolarization penalty); under OCBC the free energy is evaluated at $D = 0$, in which case $U_{dp} = [\Omega/2\varepsilon_0(1+\chi_\infty)]P^2$ is just the conventional depolarization energy.

## 3. Results

### *3.1 Three competing structural phases*

Phonon analysis of the high-symmetry $R\overline{3}c$ reference structure of $LiTaO_3$ reveals two unstable zone-center modes that are relevant to the present study. Under SCBC, the polar $A_{2u}$ mode has imaginary frequency 181$i$ cm$^{-1}$, with an eigenvector that displaces the Li and Ta cations along the +$z$ axis and the oxygen anions in the opposite direction (Fig. 1a). Condensation of this mode produces the conventional FE phase. The antipolar (or 'nonpolar' [23]) $A_{2g}$ mode has imaginary frequency 103$i$ cm$^{-1}$, with eigenvector that displaces Li atoms in adjacent unit cells in opposite directions along $z$ (Fig. 1c). Condensation of this mode produces the AFE phase, with zero net polarization at the unit-cell scale.

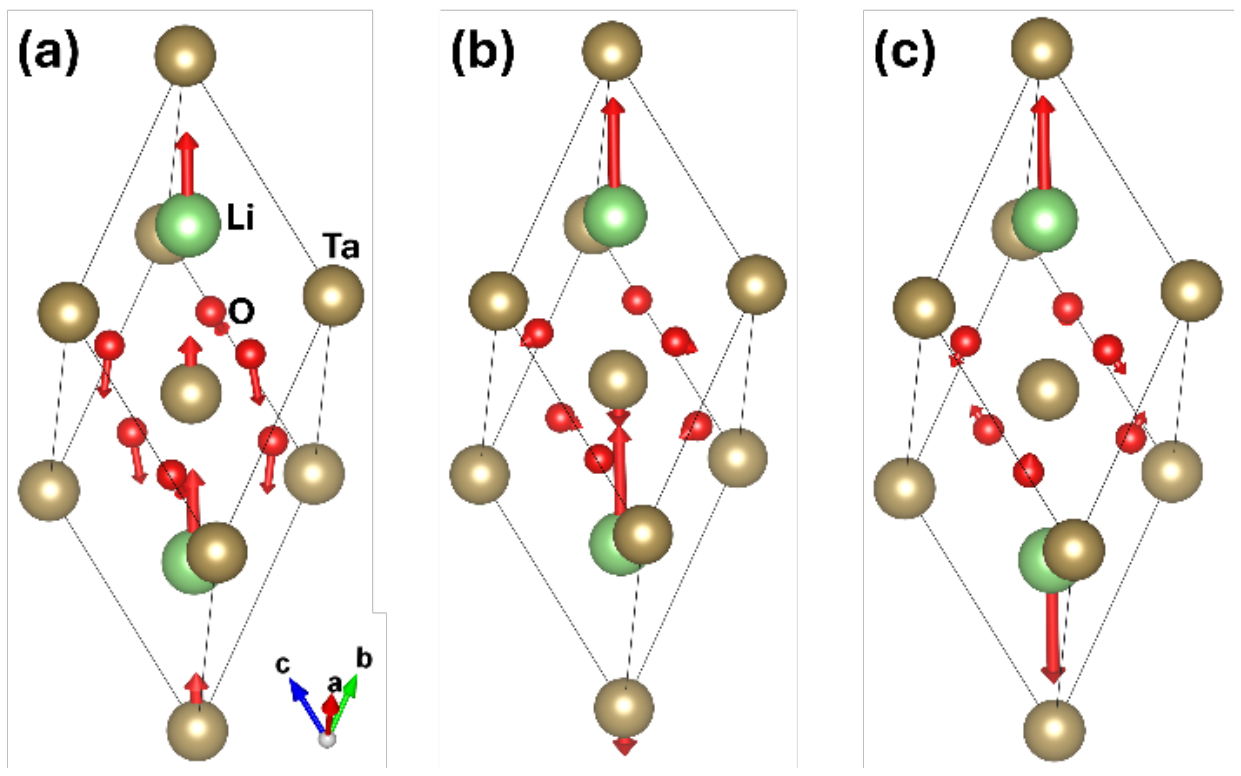

**Figure 1.** Atomic structure of $LiTaO_3$ in the high-symmetry rhombohedral unit cell. Red arrows indicate eigenvectors of (a) the soft polar $A_{2u}$ transverse-optic (TO) mode at 181$i$ cm$^{-1}$, which condenses to the FE phase; (b) the soft $A_{2u}$ longitudinal-optic (LO) mode at 48.3$i$ cm$^{-1}$, which condenses to the HyFE phase; and (c) the soft antipolar $A_{2g}$ mode at 103$i$ cm$^{-1}$, which condenses to the AFE phase. Li atoms are shown in green, Ta in gold, and O in red.

In addition, when the macroscopic depolarization field is included in the dynamical matrix, an $A_{2u}$ longitudinal-optic (LO) branch becomes weakly unstable with imaginary frequency 48.3$i$ cm$^{-1}$. Following the terminology used for the analogous $LiNbO_3$ case [21–24], we will refer to the structural phase associated with this LO mode as the hyperferroelectric (HyFE) phase (Fig. 1b). For comparison, the corresponding mode frequencies in $LiNbO_3$ are 202$i$ cm$^{-1}$ (TO), 96$i$ cm$^{-1}$ (LO), and 120$i$ cm$^{-1}$ (antipolar) [24]; the systematically lower magnitudes in $LiTaO_3$ indicate weaker structural instabilities, consistent with its lower Curie temperature.

### *3.2 Phase competition under OCBC and SCBC*

Figure 2 shows the free-energy landscapes of the three competing phases as functions of the displacement amplitude $\lambda$ under both boundary conditions. Under OCBC ($D = 0$), the free energy of the TO/FE configuration (panel a) shows that the depolarization energy $U_{dp}$ rises steeply with $\lambda$ — reaching more than 1 eV by $\lambda \approx 0.12$ — and entirely overwhelms the deep $U_{KS}$ double well that would otherwise produce ferroelectricity. The free energy $\Phi$ has its global minimum at $\lambda = 0$; the FE phase is fully destabilized.

For the LO/HyFE configuration (Fig. 2b), a double well does form in $\Phi$, but it is extraordinarily shallow: the well depth is only −0.24 meV at $\lambda \approx 0.04$. This well is two orders of magnitude shallower than the AFE well discussed below. While the LO mode is technically unstable — placing $LiTaO_3$ within the formal definition of a hyperferroelectric of Garrity, Rabe, and Vanderbilt [21] and consistent with the proposal of Li and co-workers [22] — the resulting hyperferroelectric phase is thermodynamically negligible.

For the AFE configuration (Fig. 2c), the situation is qualitatively different. Because the AFE eigenvector produces zero net unit-cell polarization, $U_{dp}$ is identically zero throughout the configuration space, and $\Phi = U$. A clear double well develops in $\Phi$ with depth −19.4 meV at $\lambda \approx 0.10$. The AFE phase is the global minimum under OCBC, more than 80 times deeper than the LO well and uncontested by either of the polar phases.

Under SCBC, the depolarization energy vanishes for all phases and $\Phi = U_{KS}$ everywhere. The polar TO phase now exhibits its full FE double well, with depth −144 meV at $\lambda \approx 0.14$ (Fig.

2d). This is far deeper than both the AFE well (−19.4 meV) and the LO well (−2.3 meV in *U*). The FE phase is therefore the unambiguous ground state under SCBC.

The two-orders-of-magnitude separation between the LO well and the AFE well at $D = 0$ justifies setting aside the LO phase from the central narrative for the remainder of the analysis. The remaining physics is the competition between the FE phase (favored at SCBC) and the AFE phase (favored at OCBC) as a function of the imposed *D* field.

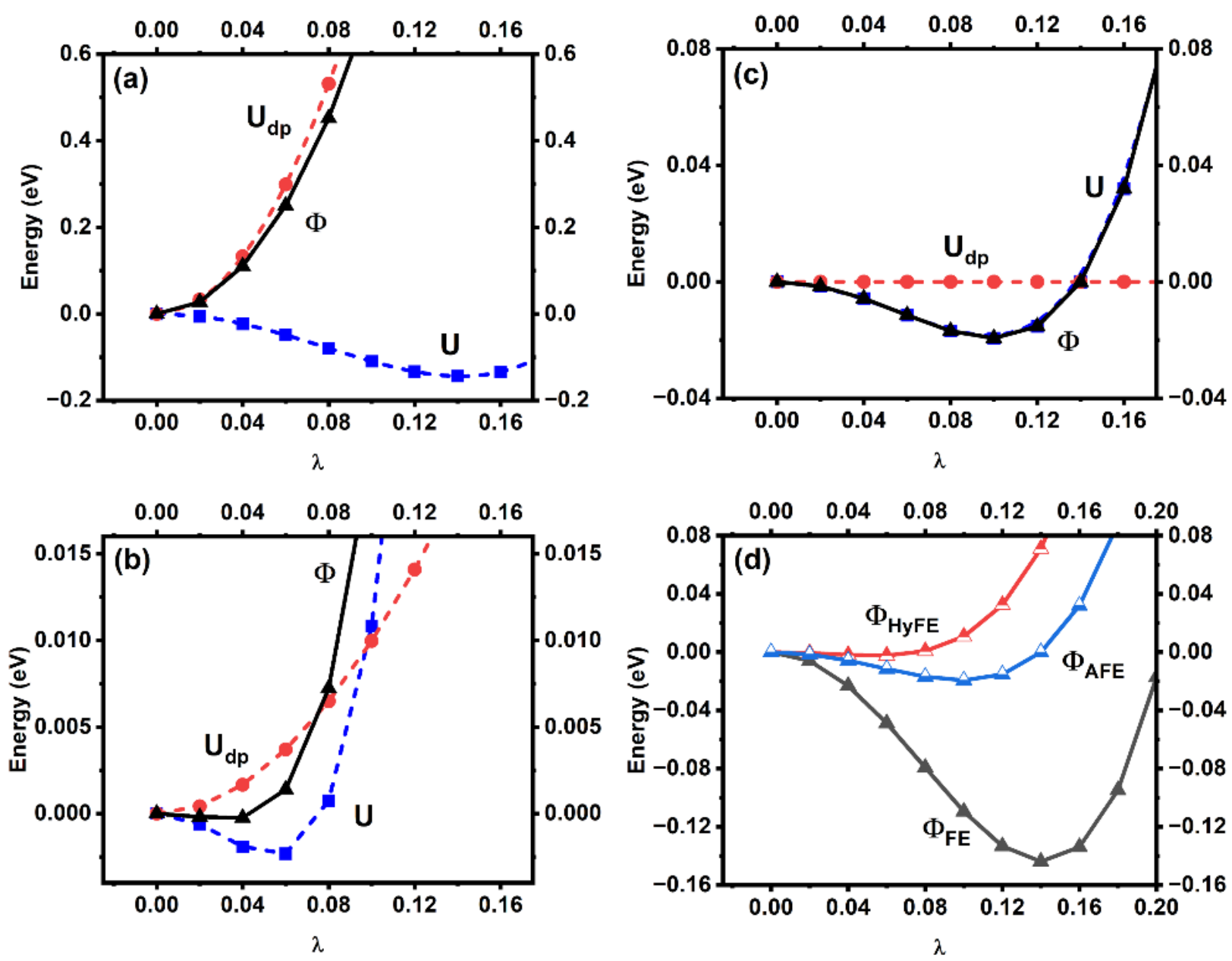

**Figure 2.** Free-energy landscapes of the three competing structural phases of $LiTaO_3$ as functions of displacement amplitude λ. Panels (a)–(c) show the TO/FE, LO/HyFE, and AFE phases under open-circuit boundary conditions (D = 0). Panel (d) compares all three phases under short-circuit boundary conditions. Internal energy $U_{KS}$ (squares), depolarization energy $U_{dp}$ (circles), and free energy Φ (triangles) are shown. Note the very different vertical scales between panels.

### *3.3 Field-induced response of the FE and AFE phases*

To understand how the relative stability of the FE and AFE phases evolves with the applied displacement field, Figure 3 compares the free-energy landscapes of the two phases at $D = 0$, 0.08, and 0.16 C/m$^2$. The TO/FE phase is shown in the left column and the AFE phase in the right column.

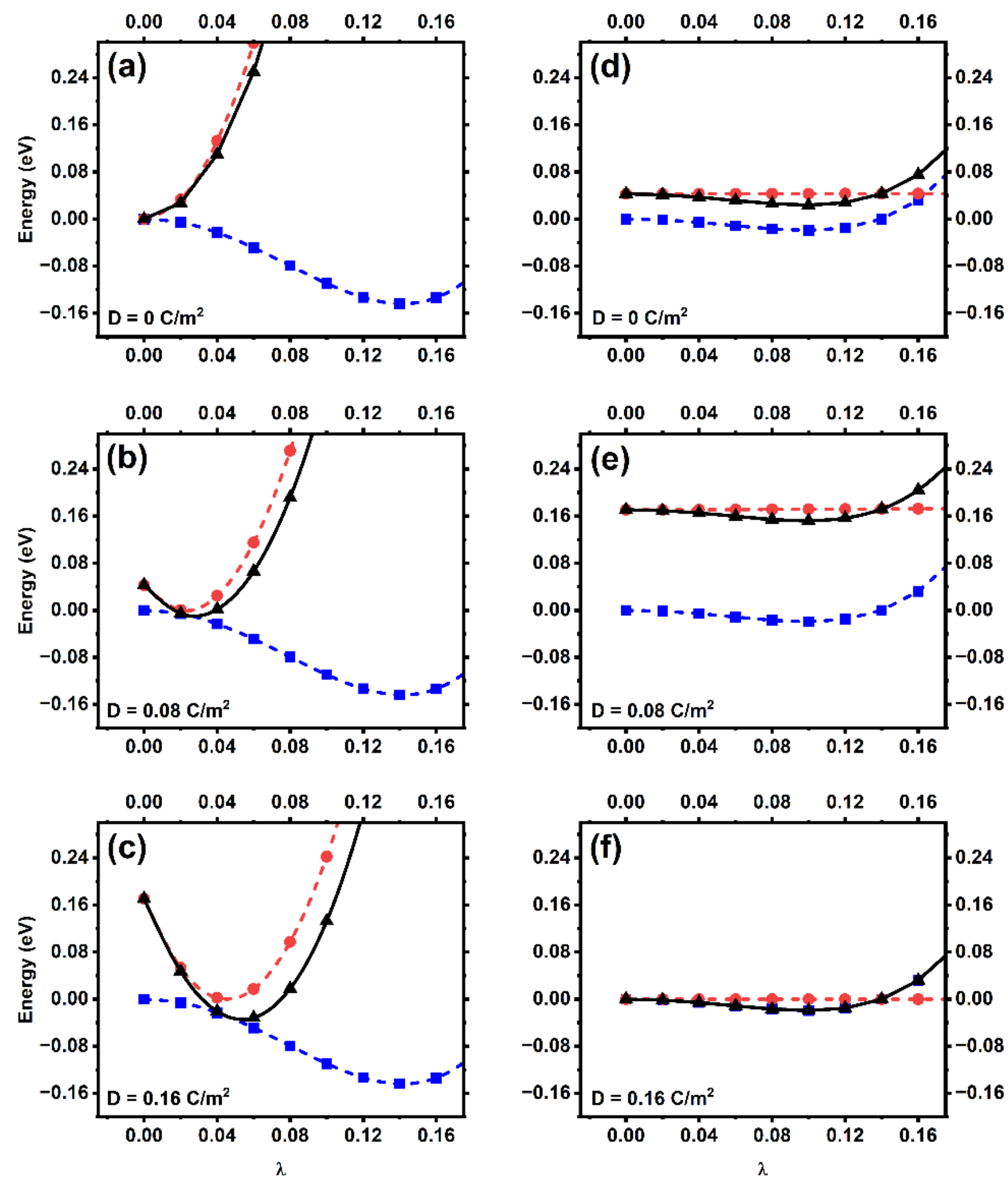

**Figure 3.** Free-energy landscapes of the FE/TO phase (left column, panels a–c) and AFE phase (right column, panels d–f) under D = 0, 0.08, and 0.16 C/m$^2$. Internal energy $U_{KS}$ (squares), depolarization energy $U_{dp}$ (circles), and free energy Φ (triangles) are shown.

For the TO/FE phase (Fig. 3a–c), the response to increasing $D$ is dramatic. At $D = 0$ the FE configuration is destabilized as already shown in Fig. 2a. As $D$ increases, the depolarization energy $U_{dp}$ — which depends on $[D - P(\lambda)]^2$ — develops a minimum at the value of λ where $P(\lambda)$ matches $D$. This shifts the global minimum of Φ to a non-zero λ. By $D = 0.16$ C/m$^2$, the FE well is well-developed at $\lambda \approx 0.06$ with depth −34.5 meV (relative to $\Phi$ at $\lambda = 0$). The FE phase progressively gains stability as $D$ increases.

The AFE response (Fig. 3d–f) is qualitatively different. Because $P(\lambda) = 0$ for the AFE eigenvector, $U_{dp}$ depends only on $D$ itself and is essentially independent of $\lambda$. Increasing $D$ therefore adds a constant offset (proportional to $D^2$) to $\Phi$ without changing its shape. The well depth — measured from $\Phi$ at $\lambda = 0$ — remains essentially constant at about 19 meV across the entire range of $D$ considered. However, the absolute value of $\Phi$ at the AFE minimum rises rapidly with $D$: at $D = 0.16$ C/m$^2$ it sits at $\Phi \approx +152$ meV in absolute terms, far above the FE minimum at the same field.

This contrast — the FE well deepens monotonically with $D$ while the AFE well remains fixed in shape but is offset upward — is the physical mechanism that produces the FE↔AFE phase transition shown in the next subsection.

### *3.4 Phase diagram and critical displacement field*

Figure 4 collects the optimal (lowest) free energy of each phase as a function of $D$. The numerical values are summarized in Table 1.

**Table 1.** Optimal free energy $\Phi_{min}$ (in meV) of the three structural phases as a function of D (in C/m$^2$). Bold indicates the ground state at each D.

| D (C/m²) | $\Phi_{FE}$ | $\Phi_{AFE}$ | $\Phi_{HyFE}$ | Ground state |
|---|---|---|---|---|
| 0.00 | 0 | **−19.4** | −0.24 | AFE |
| 0.04 | −2.0 | **−8.6** | ~−0.5 | AFE |
| 0.08 | **−10.3** | +23.5 | +16.5 | FE |
| 0.12 | **−21.0** | +76.8 | +50 | FE |
| 0.16 | **−34.5** | +152.2 | +110 | FE |

Three observations follow from Table 1 and Figure 4. First, the AFE phase is the ground state for $D \lesssim D_c$, and the FE phase is the ground state for $D \gtrsim D_c$, with crossover at $D_c$ = 0.051 C/m$^2$. Second, the HyFE phase is never the global minimum at any $D$: at small $D$ it sits between the AFE and FE phases, and at large $D$ it rises above both. The HyFE free energy values (−0.24 to −0.5 meV in the small-$D$ regime) make this point especially clearly: the HyFE phase is essentially indistinguishable from the centrosymmetric reference on the energy scale that matters for the AFE↔FE competition.

Third, the critical field $D_c$ = 0.051 C/m$^2$ in $LiTaO_3$ is appreciably lower than the corresponding value of 0.07 C/m$^2$ found for the analogous transition in $LiNbO_3$ [24]. The lower critical field follows directly from the weaker FE instability in $LiTaO_3$ (FE well 144 meV deep versus 258 meV in $LiNbO_3$): less depolarization energy is needed to overcome the polar instability and trigger the transition into the AFE state.

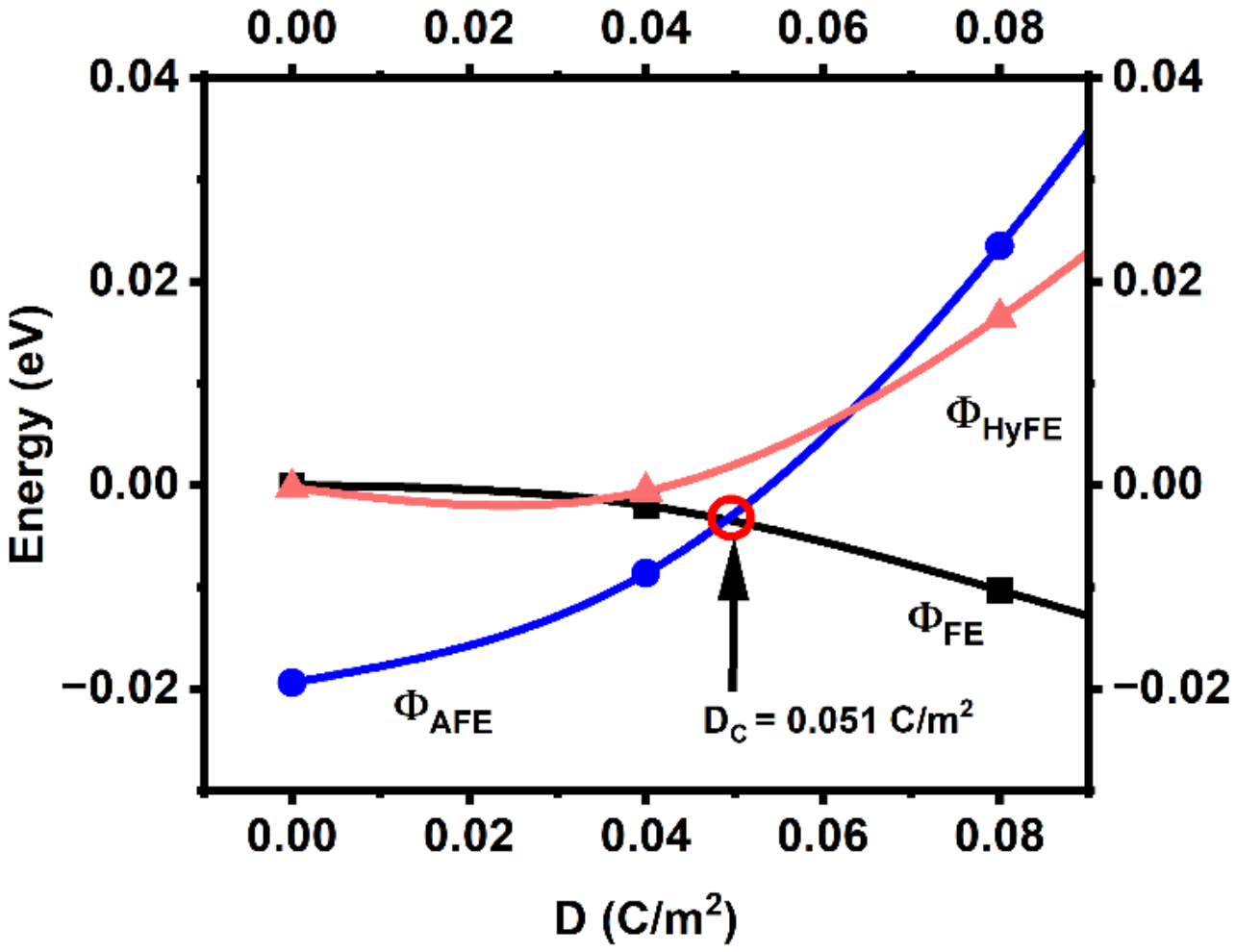

**Figure 4.** Optimal free energy of the FE, AFE, and HyFE phases as a function of electric displacement D. The crossover between the AFE and FE phases at $D_c$ = 0.051 C/m$^2$ marks the boundary-condition-driven phase transition. The HyFE curve sits between the AFE and FE curves at small D and rises above both at large D, but is never the global minimum.

### *3.5 Continuous transformation path between AFE and FE*

To examine whether the AFE↔FE transition is first or second order, we evaluate the free energy along a linear interpolation path between the AFE optimal configuration (parameter $\alpha = 0$) and the FE optimal configuration at the relevant $D$ ($\alpha = 1$). Figure 5 shows the result at $D = 0$, $D_c$ = 0.051 C/m$^2$, and $D$ = 0.08 C/m$^2$.

At $D = 0$ (Fig. 5a), the path rises monotonically from the AFE end ($\Phi = 0$) to the FE end ($\Phi \approx +20$ meV) without an intervening barrier. The depolarization energy is zero throughout because both endpoints have zero polarization (the AFE phase by symmetry, and the FE end at $D = 0$ because the FE configuration collapses to the centrosymmetric structure under OCBC). The AFE phase is the ground state with no kinetic obstacle preventing relaxation toward it.

At $D_c$ (Fig. 5b), the free-energy landscape becomes nearly flat across the entire range of $\alpha$: $\Phi$ varies only between about 10.5 meV and 18.5 meV. Notably, an intermediate local minimum appears near $\alpha \approx 0.3$ with $\Phi \approx 10.5$ meV, slightly below the AFE end ($\Phi \approx 17.4$ meV) and slightly below the FE end ($\Phi \approx 16.0$ meV). This near-degenerate, multi-minimum landscape is a direct fingerprint of phase coexistence and large fluctuations near the critical field.

At $D$ = 0.08 C/m$^2$ (Fig. 5c), the global minimum has clearly shifted to the FE end ($\alpha \approx$ 0.8–1.0, $\Phi \approx 10.4$ meV), and the AFE end now sits at $\Phi \approx +43$ meV. The path is again monotonic in the sense that no large barrier separates the AFE and FE configurations.

The smooth, continuous evolution of the free-energy landscape across these three $D$ values — with no abrupt barriers or first-order discontinuities — is consistent with a second-order phase transition, the same nature found for the analogous transition in $LiNbO_3$ [24].

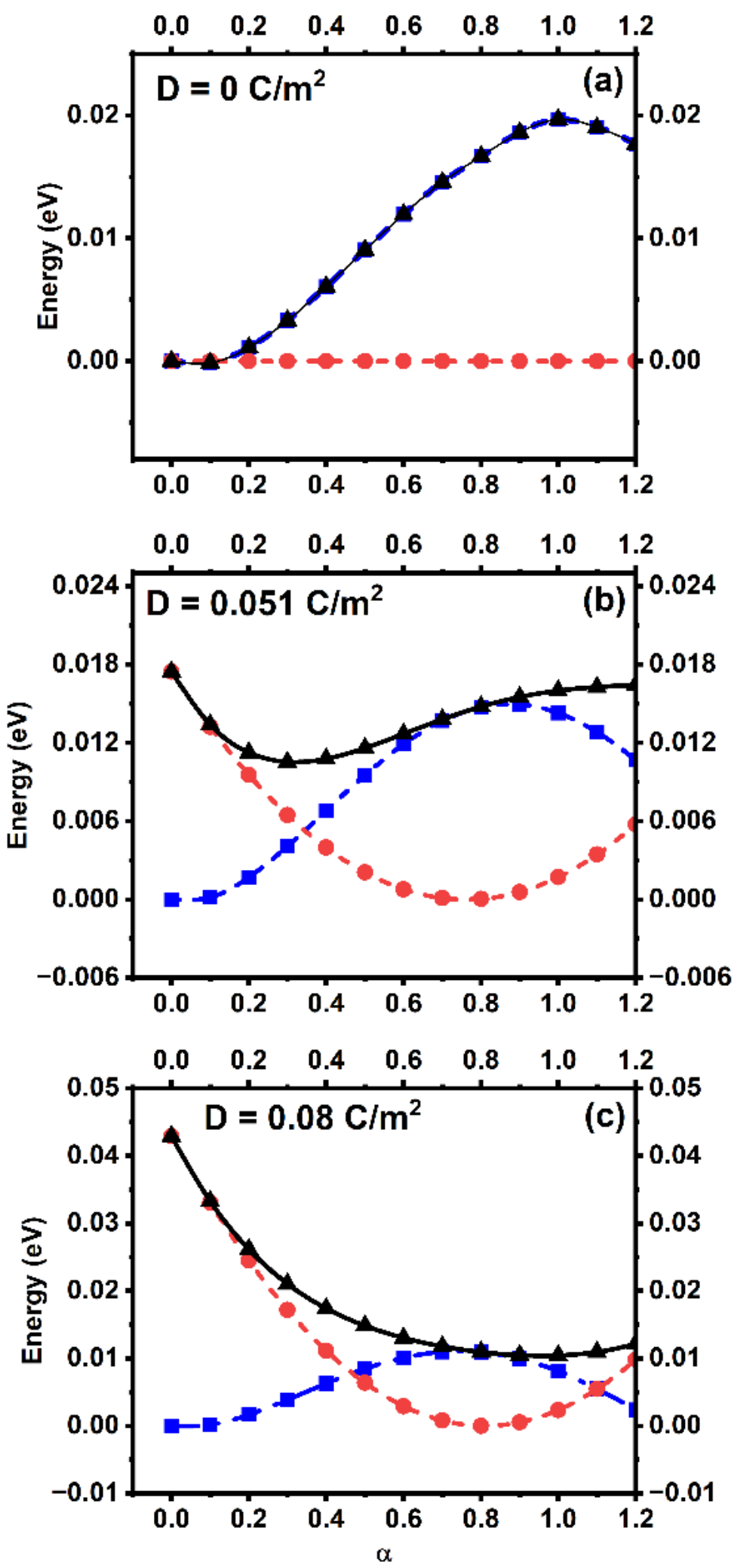

**Figure 5.** Free energy along the linear-interpolation path between the AFE configuration (α = 0) and the FE configuration (α = 1) at three values of the electric displacement: (a) D = 0, (b) $D_c$ = 0.051 C/m$^2$, and (c) D = 0.08 C/m$^2$. The flat landscape with intermediate minimum at $D_c$ is consistent with a continuous, second-order phase transition. Internal energy $U_{KS}$ (squares), depolarization energy $U_{dp}$ (circles), and free energy Φ (triangles) are shown.

## 4. Discussion

### *4.1 Comparison with $LiNbO_3$*

Our results for $LiTaO_3$ can be directly compared with the recently published study of $LiNbO_3$ under finite *D* fields [24]. Table 2 summarizes the key parameters of both materials.

**Table 2.** Comparison of $LiNbO_3$ [24] and $LiTaO_3$ (this work).

| Quantity | $LiNbO_3$ [24] | $LiTaO_3$ (this work) |
|---|---|---|
| TO mode frequency | 202i $cm^{-1}$ | 181i $cm^{-1}$ |
| LO mode frequency | 96i $cm^{-1}$ | 48.3i $cm^{-1}$ |
| Antipolar mode frequency | 120i $cm^{-1}$ | 103i $cm^{-1}$ |
| FE well depth (SCBC) | −258 meV | −144 meV |
| AFE/nonP well depth (OCBC) | −30.2 meV | −19.4 meV |
| HyFE/LO well depth (OCBC) | −13.1 meV | −0.24 meV |
| Critical field $D_c$ | 0.07 $C/m^2$ | 0.051 $C/m^2$ |
| Order of phase transition | second order | second order |

Two structural-physics conclusions follow from this comparison. First, all of the relevant indicators of FE instability are systematically weaker in $LiTaO_3$ than in $LiNbO_3$: the TO, LO and antipolar mode frequencies are all less imaginary, the FE well is roughly half as deep, and the critical *D* field is lower. This is consistent with the lower Curie temperature of $LiTaO_3$ (~950 K) compared with $LiNbO_3$ (~1480 K) [13]. The same physical mechanism — a soft polar TO mode whose freezing produces an FE ground state — operates in both materials, but with smaller magnitude in $LiTaO_3$.

Second, the difference in HyFE well depth between the two materials is striking: −13.1 meV in $LiNbO_3$ versus −0.24 meV in $LiTaO_3$, a 55-fold reduction. While $LiTaO_3$ formally satisfies the Garrity–Rabe–Vanderbilt criterion for hyperferroelectricity — it does have an unstable LO mode — the resulting phase is so weakly bound that it is thermodynamically irrelevant, and the AFE phase is the unambiguous ground state under OCBC. This refines the picture put forward by Li and co-workers [22], who identified $LiTaO_3$ as a candidate hyperferroelectric on the basis of LO-mode analysis alone. In the full free-energy comparison that includes the antipolar instability, the AFE phase decisively wins.

### *4.2 Implications for applications*

The lower critical field in $LiTaO_3$ relative to $LiNbO_3$ has practical consequences for thin-film and capacitor applications. In a thin-film geometry the effective $D$ inside the film depends on the screening efficiency of the electrodes and on the film thickness; the lower $D_c$ of $LiTaO_3$ relaxes the requirements on these parameters and brings the FE↔AFE transition into a more readily accessible experimental regime. Combined with $LiTaO_3$'s established role as a leading platform for integrated photonics [14,15], this opens routes to multifunctional devices in which the same material element can be electrically switched between an FE memory-like regime and an AFE energy-storage-like regime simply by changing the electrode boundary conditions.

The continuous, second-order character of the transition further suggests that $LiTaO_3$ near $D_c$ should exhibit large dielectric, electrocaloric, and electromechanical responses associated with the soft, near-degenerate energy landscape — a regime in which the material is structurally susceptible to small electrical or mechanical perturbations.

## 5. Conclusions

We have used first-principles density functional theory to investigate how the choice of electric boundary conditions selects among competing structural phases in $LiTaO_3$. Phonon analysis of the high-symmetry reference structure identifies three relevant unstable modes: a polar $A_{2u}$ TO mode at $181i$ cm$^{-1}$ that drives the FE phase, a weakly unstable polar LO mode at $48.3i$ cm$^{-1}$ that drives a candidate hyperferroelectric phase, and an antipolar $A_{2g}$ mode at $103i$ cm$^{-1}$ that drives the AFE phase.

Under SCBC the FE phase is the global minimum, with a free-energy well depth of −144 meV. Under OCBC the FE phase is destabilized, and the AFE phase becomes the ground state with a well depth of −19.4 meV. The HyFE phase exists but with a well depth of only −0.24 meV — nearly two orders of magnitude shallower than the AFE well — and is never the global minimum at any $D$. This refines the literature claim that $LiTaO_3$ is hyperferroelectric: it is, in the formal sense, but the magnitude of the LO instability is too small to be thermodynamically significant.

Mapping the free energy as a function of $D$ identifies a critical value $D_c = 0.051$ C/m$^2$ at which the AFE and FE phases have equal free energy. The transition is second order: the free-energy landscape along the AFE↔FE path is smooth and flat at $D_c$, with no first-order discontinuity. The lower $D_c$ in $LiTaO_3$ relative to $LiNbO_3$ (0.07 C/m$^2$) makes the transition more experimentally accessible and, combined with the technological importance of $LiTaO_3$ in integrated photonics, suggests practical routes to electrically tunable dielectrics that combine FE memory and AFE energy-storage functionality in a single material. Future work will explore the coupling between the AFE↔FE phase competition and strain, temperature, and finite-size effects in thin-film geometries.

## Acknowledgments

The support and resources from the Center for High Performance Computing at the University of Utah are gratefully acknowledged. Funds provided by the Gibson & Skaggs Research Fellowship are gratefully acknowledged.